# Protein Folding as a Quantum Transition Between Conformational States: Basic Formulas and Applications

Liaofu Luo*

Laboratory of Theoretical Biophysics,   Faculty of Physical Science and Technology,

Inner Mongolia University,   Hohhot 010021,   China

*Email address: lolfcm@mail.imu.edu.cn

## Abstract
The protein folding is regarded as a quantum transition between torsion states on polypeptide chain. The deduction of the folding rate formula in our previous studies is reviewed. The rate formula is generalized to the case of frequency variation in folding. Then the following problems about the application of the rate theory are discussed: 1) The unified theory on the two-state and multi-state protein folding is given based on the concept of quantum transition. 2) The relationship of folding and unfolding rates vs denaturant concentration is studied. 3) The temperature dependence of folding rate is deduced and the non-Arrhenius behaviors of temperature dependence are interpreted in a natural way. 4) The inertial moment dependence of folding rate is calculated based on the model of dynamical contact order and consistent results are obtained by comparison with one-hundred-protein experimental dataset. 5) The exergonic and endergonic foldings are distinguished through the comparison between theoretical and experimental rates for each protein. The ultrafast folding problem is viewed from the point of quantum folding theory and a new folding speed limit is deduced from quantum uncertainty relation. And finally, 6) since only the torsion-accessible states are manageable in the present formulation of quantum transition how the set of torsion-accessible states can be expanded by using statistical energy landscape approach is discussed. All above discussions support the view that the protein folding is essentially a quantum transition between conformational states.

# 1 Basic theory – a brief review of the deduction

We have deduced protein folding rate from the quantum transition between vibration states of different conformations. Since the transition is generally related to the electronic states of the molecule we suppose the dynamical variables of the system are ($\theta, x$) ($x$ is the coordinate of the frontier electron and θ the torsion angle of molecule). The wave function of the system $M(\theta, x)$ satisfies

$$(H_1(\theta, \frac{\partial}{\partial \theta}) + H_2(\theta, x, \nabla))M(\theta, x) = EM(\theta, x) \tag{1}$$

Starting from (1) the main steps of the deduction are summarized in the following. The detailed discussions were given in literatures [Luo 1987; Luo 1995; Luo 2004; Luo 2009].

**(1) Adiabatic approximation** Because the mass of electrons is much smaller than nucleic masses, the adiabatic approximation can be used. In this approximation the wave function of the conformation-electronic system can be expressed as

$$M(\theta, x) = \psi(\theta)\varphi(x, \theta) \tag{2}$$

and these two factors satisfy

$$H_2(\theta, x, \nabla)\varphi_\alpha(x, \theta) = \varepsilon^\alpha(\theta)\varphi_\alpha(x, \theta) \tag{3}$$

$$\{H_1(\theta, \frac{\partial}{\partial \theta}) + \varepsilon^\alpha(\theta)\}\psi_{kn\alpha}(\theta) = E_{kn\alpha}\psi_{kn\alpha}(\theta) \tag{4}$$

here $\alpha$ denotes the electron-state, and ($k, n$) refer to the conformation- and vibration-state, respectively.

**(2) Quantum transition by use of the nonadiabaticity operator method** [Huang and Rhys, 1950]. Because Eq (2) is not a rigorous eigenstate of Hamiltonian $H_1 + H_2$, there exist transitions between adiabatic states that result from the off–diagonal elements

$$\int M^+_{k'n'\alpha'}(H_1 + H_2)M_{kn\alpha}d\theta d^3x = E_{kn\alpha}\delta_{kk'}\delta_{nn'}\delta_{\alpha\alpha'} + \langle k'n'\alpha' | H' | kn\alpha \rangle \tag{5}$$

$$\langle k'n'\alpha' | H' | kn\alpha \rangle = \int \psi^+_{k'n'\alpha'}(\theta)\{-\frac{\hbar^2}{2I}\int \varphi^+_\alpha(\frac{\partial^2 \varphi_\alpha}{\partial \theta^2} + 2\frac{\partial \varphi_\alpha}{\partial \theta}\frac{\partial}{\partial \theta})d^3x\}\psi_{kn\alpha}(\theta)d\theta \tag{6}$$

Here $H'$ is a Hamiltonian describing conformational transition. We see that the conformational transition is caused by the matrix element of conformational wave function $\psi_{kn\alpha}(\theta)$ but partly related to the electronic wave function $\varphi_a(x, \theta)$ through its $\theta$ dependence.

For most protein folding problem the electronic state does not change in transition processes, namely $\alpha' = \alpha$. Because the wave function $\varphi_a$ is generally real, one can deduce

$$\int \varphi_a(x, \theta)\frac{\partial \varphi_a(x, \theta)}{\partial \theta}d^3x = 0 \tag{7}$$

from the normalization condition $\int \varphi_a(x, \theta)\varphi_a(x, \theta)d^3x = 1$. Therefore, only the first term in Eq. (6) should be retained, namely

$$\langle k'n'\alpha | H' | kn\alpha \rangle = \int \psi^+_{k'n'\alpha}(\theta)\{-\frac{\hbar^2}{2I}\int \varphi^+_\alpha \frac{\partial^2 \varphi_\alpha}{\partial \theta^2}d^3x\}\psi_{kn\alpha}(\theta)d\theta \tag{8}$$

**(3) Application to protein folding**   For a problem of fixed bond lengths and bond angles only the torsion potential $U_{tor}(\theta_1,...\theta_N)$ should be considered. The torsion potential $U_{tor}$ is assumed to have several minima with respect to each $\theta_i$ and near each minimum the potential can be expressed by a potential of harmonic oscillator.   In argument-separable case

$$U_{tor}(\theta_1,\theta_2,...\theta_N) = \sum_j U_{tor}^{(j)}(\theta_j) \tag{9}$$

we have

$$H_1 = \sum (-\frac{\hbar^2}{2I_j}\frac{\partial^2}{\partial \theta_j^2} + U_{tor}^{(j)}(\theta_j)) \tag{10}$$

here $I_j$ is the inertial moment of the *j*-th mode. The conformational wave function is the product of the functions of single argument

$$\psi_{kn\alpha}(\theta) = \psi_{k_1,n_1,\alpha_1}(\theta_1)......\psi_{k_N,n_N,\alpha_N}(\theta_N) \tag{11}$$

where $\psi_{k_j,n_j,\alpha_j}(\theta_j)$ can approximately be expressed by a wave function of harmonic oscillator with quantum number $n_j$. Note that the harmonic potential has equilibrium position at $\theta_j = \theta_{k_j}^{(0)}$ with the corresponding $k_j$-th minimum of potential $E_{k_j}$ ($k_j$=1,2,…).

Finally, by use of the nonadiabaticity operator method and by consideration of $\alpha' = \alpha$ in the process, like Eq (8), we obtain the conformation-transitional matrix element

$$\langle k'n'\alpha | H' | kn\alpha \rangle = \int \psi_{k'n'\alpha}^+(\theta) \sum \{-\frac{\hbar^2}{2I_j}\int \varphi_\alpha^+ \frac{\partial^2 \varphi_\alpha}{\partial \theta_j^2} d^3x\} \psi_{kn\alpha}(\theta)d\theta \tag{12}$$

**(4) Deduction of transition rate – single mode case**

After thermal average over the initial states and summation over final states we have

$$W = \frac{2\pi}{\hbar}\sum_{\{n\}}|\langle k'n'\alpha | H' | kn\alpha \rangle|^2 B(\{n\},T)\rho_E \tag{13}$$

$B(\{n\},T)$ denotes the Boltzmann factor and $\rho_E$ means state density. The final result of transition rate for single mode case is given by

$$W = \frac{2\pi}{\hbar^2 \omega}I_E I_V \tag{14}$$

$$I_E = \left|\frac{-\hbar^2}{2I}\int \varphi_\alpha \frac{\partial^2}{\partial \theta^2}\varphi_\alpha d^3x\right|^2_{\theta=\theta_0} \tag{15}$$

$$I_V = (2\pi z)^{-1/2}\exp(-\frac{p^2}{2z})\exp\frac{\delta E}{2k_B T} \tag{16}$$

$$(z = (\delta\theta^2)\frac{k_B T}{\hbar^2}I, \quad p = \frac{\delta E}{\hbar\omega}) \qquad (17)$$

Here the same frequency for initial and final states ($\omega=\omega'$) has been assumed, $\theta_0$ means the conformational coordinate which takes a value of the largest overlap region of vibration functions, and $\delta\theta = \theta_k^{(0)} - \theta_{k'}^{(0)}$ is the angular displacement and $\delta E = E_k - E_{k'}$ the energy gap between initial and final states ($E_k$ and $E_{k'}$ mean the minimum of the potential in initial state k and final state k' respectively) (Fig 1 and Fig 2). The similar formula of Eq (16) was obtained by Jortner in his calculation of activation energy for electron transfer between biological molecules [Jortner, 1976; Devault, 1980].

### (5) Deduction of transition rate – multi- mode case

If several torsion angles participate simultaneously in one step of conformational transition then we call it multi-mode or multi-torsion transition. Consider the case of $N$ modes with same frequency $\omega_1 = \omega_2 = ... = \omega_N = \omega$, $\omega'_1 = \omega'_2 = ... = \omega'_N = \omega'$, and $\omega=\omega'$. We obtain

$$W = \frac{2\pi}{\hbar^2\omega}I_E \sum_{\{p_j\}}\prod_j I_{Vj} \qquad (18)$$

$I_E$ is the factor of electronic wave function

$$I_E = \left|\sum_j (\frac{-\hbar^2}{2I_j}\int \varphi_\alpha \frac{\partial^2}{\partial\theta_j^2}\varphi_\alpha d^3x)\big|_{\theta_j=\theta_{j0}}\right|^2 \qquad (19)$$

and $\sum_{\{p_j\}}\prod_j I_{Vj}$ is the factor of conformational wave function

$$\sum_{\{p_j\}}\prod_j I_{Vj} = \frac{1}{\sqrt{2\pi}}\exp(\frac{\Delta E}{2k_B T})(\sum Z_j)^{-\frac{1}{2}}\exp(-\frac{p^2}{2\sum Z_j}) \qquad (20)$$

$$(\sum_j p_j = p = \Delta E/\hbar\omega \qquad \Delta E = \sum \delta E_j, \quad Z_j = (\delta\theta_j^2)\frac{k_B T}{\hbar^2}I_j) \qquad (21)$$

(Here $\delta\theta_j$ is the angular displacement and $\delta E_j$ the energy gap between the initial and final states for the $j$-th mode.)

### (6) Simplified results
By using

$$I_E = \left\{ \sum \frac{\hbar^2}{2I_j} \langle l_j^2 \rangle \right\}^2$$

$$= \frac{\hbar^4}{4} \left( \sum_j \frac{a_j}{I_j} \right)^2 \quad a_j = \langle l_j^2 \rangle \qquad (22)$$

$l_j$ = the $j$-th magnetic quantum number (with respect to $\theta_j$) of electronic wave function $\varphi_\alpha(x, \{\theta\})$, $a_j = <l_j^2> \approx O(1)$, a number in the order of magnitude of 1, we obtain

$$W = \frac{\hbar^3 \sqrt{\pi}}{2\sqrt{2}\delta\theta\omega} (k_B T)^{-1/2} \exp\{\frac{\Delta E}{2k_B T}\} (\sum I_j)^{-1/2} (\sum \frac{a_j}{I_j})^2 \exp\{\frac{-(\Delta E)^2}{2\omega^2 (\delta\theta)^2 k_B T \sum I_j}\}$$
(23)

$$(\delta\theta = \sqrt{<(\delta\theta_j)^2>_{av}}) \qquad (24)$$

from Eqs (18)-(22). Setting $\omega = 6 \times 10^{12} \sec^{-1}$ (the typical frequency of torsion vibration), $\delta\theta = \pi/2$ and $T_0$=310 K ($k_B T_0 / \hbar\omega = 6.8$), neglecting the last factor in Eq (23), finally we deduce

$$W = 0.37 \times 10^{-87} (\frac{T}{T_0})^{-1/2} \exp(\frac{\delta E}{2k_B T}) a^2 I^{-2.5} \sec^{-1} \qquad (25)$$

($a = \langle l^2 \rangle \simeq O(1)$, and $I$ in unit of [g cm$^2$]) for single torsion transition and

$$W = 0.37 \times 10^{-87} (\frac{T}{T_0})^{-1/2} \exp\frac{\Delta E}{2k_B T} (\sum I_j)^{-1/2} (\sum \frac{a_j}{I_j})^2 \sec^{-1} \qquad (26)$$

($a_j \simeq O(1)$, $I_j$ in unit of [g cm$^2$]) for multi-mode case.

**(7) Generalization to frequency variation case**
Conside the case of non-equal frequencies between initial and final states and among different modes. Following classical theory of reaction rate the rate is generally a function of free energy difference between initial and final states. For a system of oscillators, the free energy is expressed by

$$G_{os} = \frac{1}{\beta} \sum_j \{\ln(1 - \exp(-\beta\omega_j)) + E_j\} \quad (\beta = \frac{1}{k_B T}) \qquad (27)$$

from statistical mechanics where $E_j$ means the potential minimum of the $j$-th oscillator. By using

$$\frac{\partial G_{os}}{\partial (\hbar \omega_j)} = \langle n_j \rangle = \frac{1}{\exp(\beta \hbar \omega_j) - 1} \tag{28}$$

($n_j$ – phonon number of the j-th mode) we obtain the free energy variation

$$\delta_\omega G = \sum_j \int_{\omega_j}^{\omega'_j} \frac{\partial G_{os}}{\partial \omega_j} d\omega_j = \sum_j \int_{\omega_j}^{\omega'_j} \frac{\hbar}{\exp(\beta \hbar \omega_j) - 1} d\omega_j \tag{29}$$

as the frequency shifts from $\omega_j$ to $\omega_j'$. It leads to

$$\delta_\omega G = \sum_j \frac{1}{\beta} \ln \frac{\omega_j'}{\omega_j} \tag{30}$$

as $\beta \hbar \omega_j \ll 1$. Therefore, from (27) the free energy difference between torsion initial state (frequency $\{\omega_j\}$) and final state (frequency $\{\omega_j'\}$) is

$$\Delta G_{os} = -\delta_\omega G + \sum_j \delta E_j = \Delta E + \sum_j \frac{1}{\beta} \ln \frac{\omega_j}{\omega_j'} \tag{31}$$

Correspondingly, the net change of phonon number $p$ in Eqs (17) and (21) will be replaced by

$$p = \sum_j \frac{\delta E_j}{\hbar \omega_j} + \sum_j \frac{1}{\beta} \frac{\ln \frac{\omega_j}{\omega_j'}}{\hbar \omega_j} \tag{32}$$

Considering that the contribution of frequency variation to folding rate comes mainly from the torsion vibration term, instead of Eq (23) we obtain a generalized equation of folding rate

$$W = \frac{\hbar^3 \sqrt{\pi}}{2\sqrt{2} \delta \theta \overline{\omega}'} (k_B T)^{-1/2} \exp\{\frac{\Delta G_{os}}{2 k_B T}\} (\sum I_j)^{-1/2} (\sum \frac{a_j}{I_j})^2 \exp\{\frac{-(\Delta G_{os})^2}{2 \overline{\omega}^2 (\delta \theta)^2 k_B T \sum I_j}\} \tag{33}$$

($\overline{\omega}'$ means the average of $\omega_j'$ and $\overline{\omega}$ -- the average of $\omega_j$) in the frequency variation case where $\Delta G_{os}$ is given by Eq (31). The application of free energy gap $\Delta G_{os}$ instead of energy gap $\Delta E$ was firstly proposed in the photosynthetic electron transfer when the frequency variation is small [Kakitani, 1981]. From Eq (31) we know that when

$$\sum_j \ln \frac{\omega_j}{\omega_j'} \ll \frac{\Delta E}{k_B T} \tag{34}$$

the effect of frequency variation can be neglected and Eq (33) returns to Eq (23).

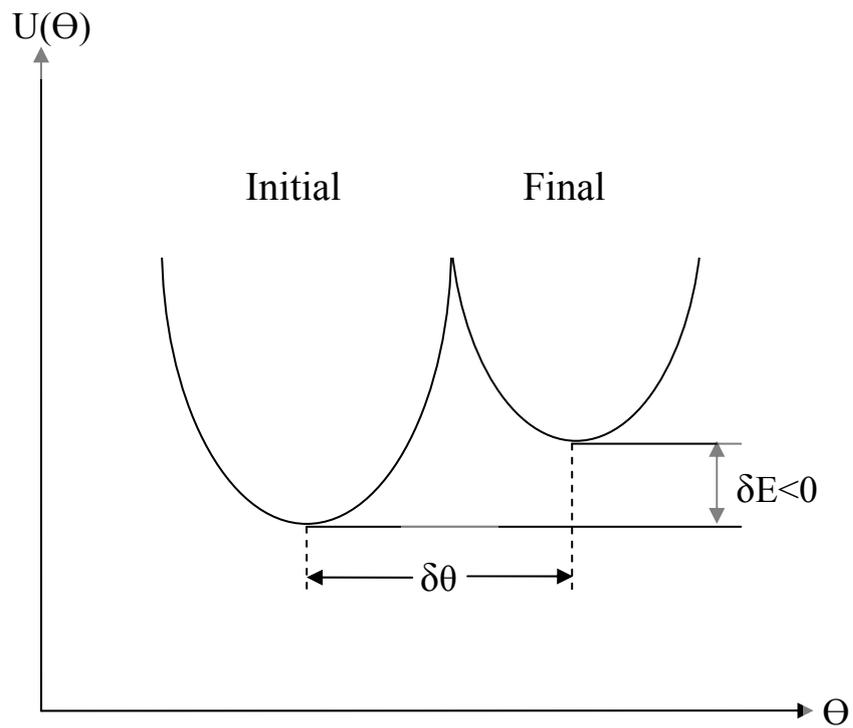

Figure 1  U(θ) – θ relation for δE<0

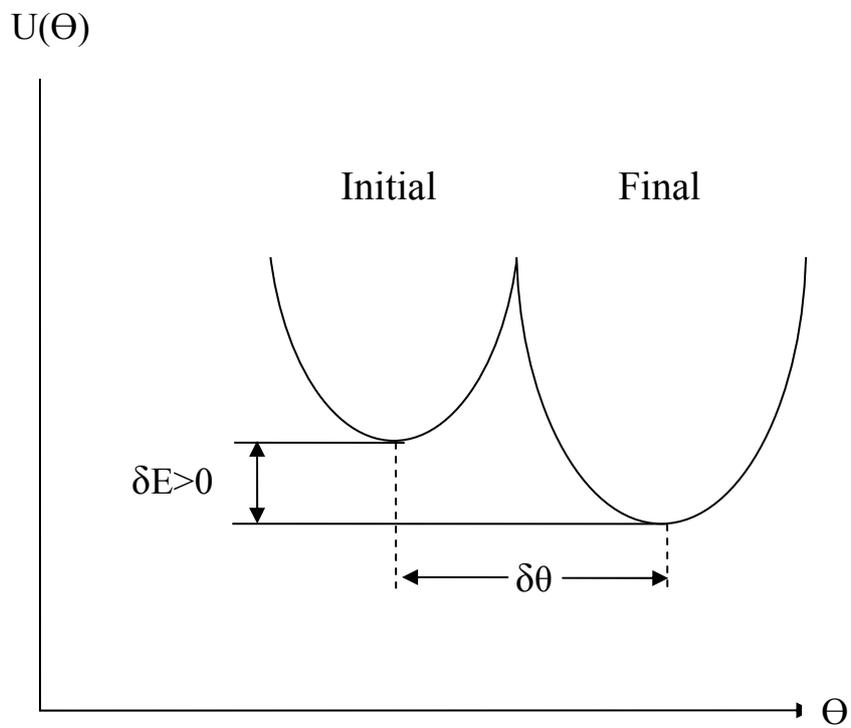

Figure 2  U(θ) – θ relation for δE>0

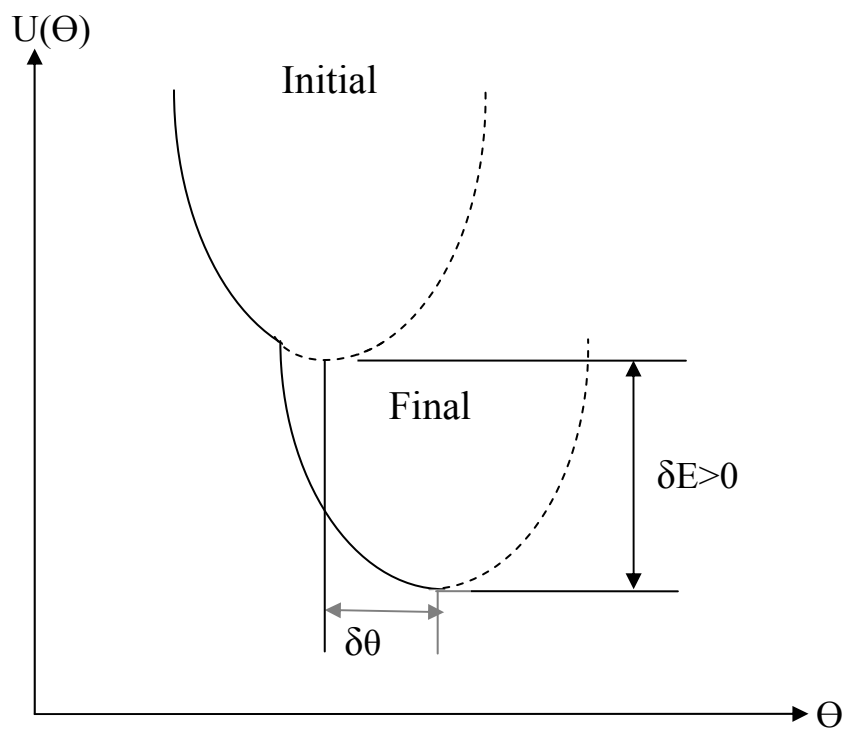

Figure 3    U(θ) – θ relation for down-hill folding

**Figure caption**    Figure 1 to 3 are plotted for typical torsion mode $j$. The subscript $j$ in $\theta_j$ and $E_j$ has been dropped for simplicity. Figure 3 is similar to Figure 2 with δE>0 but the barrier disappears in Fig 3.

## 2   Two-state and multi-state proteins

The folding kinetics of dozens of proteins have been reported to date [Fulton et al, 2005]. For two-state proteins the folding kinetics can be directly understood in the scheme of quantum transition from the denatured state to the folding state. In this section we shall generalized the theory to the multi-state protein. Since the folding rates of non-two-state proteins show the similar dependence on the native backbone topological parameters as for the two-state proteins [Kamagata et al, 2004], we assume that the multi-state protein folding can be seen as a joint process of several quantum transitions, and they occur at different degrees of freedom of torsion angle. Consider three-state (denatured state, intermediate state and folding state) case. There exist two quantum transitions, one from denatured to intermediate, referred to as transition 1 and the other from intermediate to folding state, referred to as transition 2. We suppose that the two successive transitions ($s=1,2$) in three-state proteins take place in different torsion modes, i.e., the oscillator modes in transition 1 not overlapped with those in transition 2.

For simplicity consider the case of single frequency $\omega_1 = \omega_2 = ... = \omega_N = \omega$, $\omega'_1 = \omega'_2 = ... = \omega'_N = \omega'$ and $\omega=\omega'$. Set $p_j^{(s)}$ representing the net change in quantum number for oscillator mode $j$ in transition $s$ and satisfying the constraint

$$\sum_j p_j^{(s)} = p^{(s)} = \Delta E^{(s)}/\hbar\omega \qquad \Delta E^{(s)} = \sum_j \delta E_j^{(s)} \qquad (35)$$

where $\delta E_j^{(s)}$ means the energy gap between the initial and final states for the $j$-th mode in transition $s$. Following above assumption the total oscillator modes in two successive transitions are fully identical with those in some single transition. So we have

$$\sum_{\Delta E^{(1)}} \sum_{\{p_j^{(1)}\}} \sum_{\{p_{j'}^{(2)}\}} \prod_j I_{Vj}^{(1)} \prod_{j'} I_{Vj'}^{(2)} = \sum_{\{p_j\}} \prod_j I_{Vj} \qquad (36)$$

where the constraints (35) and $\Delta E^{(1)} + \Delta E^{(2)} = \Delta E$ should be satisfied in the left summation of Eq (36). The RHS of Eq (36) which describes a single transition has been given by Eq (20). The factor $\sum_{\{p_j^{(1)}\}} \prod_j I_{Vj}^{(1)}$ or $\sum_{\{p_j^{(2)}\}} \prod_j I_{Vj}^{(2)}$ in LHS of Eq (36) takes the same approximate expression like Eq (20). Eq (36) means the conformational factor of three-state protein folding rate is equal to that of an equivalent two-state protein folding.

The electronic factor of three-state protein folding rate in transition $s$ ($s=1,2$) is

$$I_E^{(s)} = \left| \sum_{j \text{ in} \{s\}} \left( \frac{-\hbar^2}{2I_j} \int \varphi_\alpha \frac{\partial^2}{\partial \theta_j^2} \varphi_\alpha d^3 x \right) \right|_{\theta_j=\theta_{j0}} \Bigg|^2$$

$$\cong \frac{\hbar^4}{4}(\sum_{j \text{ in}\{s\}} \frac{a_j}{I_j})^2 \tag{37}$$

Comparing with Eq (22) we obtain

$$\sqrt{I_E^{(1)}} + \sqrt{I_E^{(2)}} = \sqrt{I_E} \tag{38}$$

and therefore estimate

$$I_E^{(1)} \approx (\frac{N_1}{N_1 + N_2})^2 I_E \qquad I_E^{(2)} \approx (\frac{N_2}{N_1 + N_2})^2 I_E \tag{39}$$

where $N_1$ and $N_2$ are number of modes in transition 1 and 2 respectively. The electronic factor of three-state folding rate is the geometrical average of $I_E^{(1)}$ and $I_E^{(2)}$, $\sqrt{I_E^{(1)} I_E^{(2)}} = \frac{N_1 N_2}{(N_1 + N_2)^2} I_E$. Therefore, for three-state protein the electronic factor of folding rate is smaller than that in an equivalent two-state folding by $\frac{N_1 N_2}{(N_1 + N_2)^2}$. Combining with the conformational factor Eq (36) we estimate that the three-state protein folding rate is about $\left\langle \frac{N_1 N_2}{(N_1 + N_2)^2} \right\rangle$ (smaller than 1/4) of the corresponding equivalent two-state protein folding. The above discussion can be generalized to any multi-state protein folding. For $s$-state protein, the folding rate is multiplied by a factor smaller than $1/(s-1)^2$ as compared with the two-state protein of same topology. If there exists additional time delay $\tau$' between two successive transitions the folding rate of an $s$-state protein should further lower down by an additional factor $\exp(-(s-2)\tau')$.

To summarize, under the assumption of the successive transitions of a multi-state proteins take place in different torsion modes the multi-state protein folding rate can be simply expressed as the rate of two-state protein folding of same topology multiplied by a time delay factor $\exp(-\tau)$ ($\tau$ depends on $s$). Therefore, Eq (33) should be generalized to

$$W = \frac{\hbar^3 \sqrt{\pi}}{2\sqrt{2}\delta\theta\overline{\omega}'}(k_B T)^{-1/2} \exp\{\frac{\Delta G_{os}}{2k_B T}\}(\sum I_j)^{-1/2}(\sum \frac{a_j}{I_j})^2 \exp\{\frac{-(\Delta G_{os})^2}{2\overline{\omega}^2(\delta\theta)^2 k_B T \sum I_j}\} \exp(-\tau) \tag{40}$$

or its simplified version Eq (26) generalized to

$$W = 0.37 \times 10^{-87} (\frac{T}{T_0})^{-1/2} \exp\frac{\Delta E}{2k_B T}(\sum I_j)^{-1/2}(\sum \frac{a_j}{I_j})^2 \exp(-\tau) \sec^{-1} \tag{41}$$

for multi-state protein folding. The existence of time delay factor has been proved by Zhang and Luo through the unified statistical analysis of two-state and non-two-state protein folding by use of Eqs (26) and (41) [Zhang and Luo, 2010] . For a set of 80 proteins (45 two-state and 35 non-two-state) they estimated the best-fit $\tau = 3.5$ (see section 6).

## 3  Denaturant dependence of folding and unfolding rates

Consider a transition process (folding) from conformation A to B. The transition rate is supposed to be $W_{AB}$. For the inverse process (unfolding) the rate is $W_{BA}$. From the above general deduction of the rate equation we see that two processes are related by $\delta E_j \to -\delta E_j, \delta \theta_j \to -\delta \theta_j, p_j \to -p_j$ and $\omega_j \leftrightarrow \omega_j'$. So from (33), to the first order of $\Delta G_{os}$ one has

$$W_{AB}/W_{BA} \cong \exp(\Delta G_{os}/k_B T)\frac{\overline{\omega}}{\overline{\omega}'} \tag{42}$$

In the majority of folding kinetics experiments, chemical denaturants are employed to destabilize the native state. How the folding and unfolding rates depend on denaturant concentration? There are at least two factors influencing the folding kinetics. The first is the change of the energy gap $\delta E_j$ due to the alteration of torsion energy $U_{tor}^{(j)}(\theta_j)$ of the $j$-th mode. One may assume $\delta E_j$, and therefore $\Delta E$, is a linear function of denaturant concentration. This interprets the observed slope of the arm of the chevron plot, a plot of $\ln W$ versus denaturant concentration [Maxwell et al, 2005]. The second is the change of the inertial moment $I_j$ due to the varying viscosity of the solvent. One may assume the effective inertial moment $I_j$ is approximately a linear function of denaturant concentration. This can interpret the observed deceleration of folding as the viscosity of the solvent is increased [Jacob et al, 1997; Pabit et al, 2004]. Inserting the denaturant-dependent factor into the simplified equation (26) we obtain

$$\ln W = \ln W_0 + \frac{m_E}{2k_B T}(c - c_0) + \sum_j \frac{\partial F}{\partial I_j} m_I (c - c_0) \tag{43}$$

with

$$F(I_1,...,I_j,...) = -\frac{1}{2}\ln(\sum_j I_j) + 2\ln(\sum_j \frac{a_j}{I_j}) \tag{44}$$

where $c$ means denaturant concentration and $\ln W_0$ – the logarithm rate at standard concentration $c_0$, and $m_E$, $m_I$ denote the slop of $\Delta E$ and $I_j$ with concentration respectively. For $I_j = I$, Eq (43) is approximated as

$$\ln W = \ln W_0 + \frac{m_E}{2k_B T}(c - c_0) - \frac{5}{2}\frac{m_I}{I}(c - c_0) \qquad (45)$$

## 4 Temperature dependence of transition rate

From (33) and (31) we obtain the explicit temperature dependence of $W$

$$W \sim \exp\{\frac{\Delta G_{os}(1 - \frac{\Delta G_{os}}{\varepsilon})}{2k_B T}\}T^{-\frac{1}{2}} \qquad (46)$$

$$(\varepsilon = \bar{\omega}^2 (\delta\theta)^2 \sum I_j) \qquad (47)$$

where $\Delta G_{os}$ is temperature-dependent as seen from Eq (31). The dependent factor $T^{-\frac{1}{2}}$ has a simple classical-mechanical explanation. The conformational transition can be looked as a set of oscillators jumping across a barrier. The oscillators are in thermal equilibrium, obeying normal distribution. The factor $T^{-\frac{1}{2}}$ comes from the standard deviation of the oscillator displacement.

From Eq (46) it gives

$$\ln W(T) \sim const. + \frac{\Delta E(1 - \frac{\Delta E}{\varepsilon})}{2k_B T} + \frac{1}{2}\ln\frac{1}{T}$$

$$+ \frac{1}{2}\lambda(1 - \frac{\Delta E}{\varepsilon}) - \frac{\lambda^2}{2\varepsilon}k_B T \qquad (48)$$

$$(\lambda = \sum_j \ln\frac{\omega_j}{\omega_j'}) \qquad (49)$$

So，the relation of $\ln W$ versus $\frac{1}{T}$ is not linear in the Arrhenius plot. It behaves approximately as a straight line with slop

$$S = \frac{\Delta E(1 - \frac{\Delta E}{\varepsilon})}{2k_B} \qquad (50)$$

at high $\frac{1}{T}$, but the slop changes significantly at low $\frac{1}{T}$. The rigorous slop – temperature relation is given by

$$\frac{d\ln W}{d(\frac{1}{T})} = \frac{\Delta E(1-\frac{\Delta E}{\varepsilon})}{2k_B} + \frac{1}{2}T + \frac{k_B\lambda^2}{2\varepsilon}T^2 \qquad (51)$$

As shown in Eq (50), the symbol of $S$ is determined by the energy gap $\Delta E$. In view of $\Delta E$ generally smaller than $\varepsilon$, as $\Delta E <0$, the rate increases with temperature, across a maximum and then decreases. As $\Delta E >0$, no such maximum can be observed but the slop $S$ of Arrhenius line also increases in high temperature region.

The nonlinearity of $\ln W - \frac{1}{T}$ relation comes from the last two terms of the RHS of Eq (51). Both two terms are positive and they interpret the observed nonlinearity in the correct direction. However, the contribution of $\frac{1}{2}T$ term alone is not enough for interpreting the experimental strong nonlinearity [Yang and Gruebele, 2004; Ghosh et al, 2007]. This can be seen from the following observation. From Eqs (50) and (51) the slop $S$ of Arrhenius line is about $\frac{\Delta E}{2k_B} \sim 2T_0 = 620$ while the $\frac{1}{2}T$ term contributes to the slope increase only of 5/620≈1% when temperature changes 10°K. Therefore, the $\frac{k_B\lambda^2}{2\varepsilon}T^2$ term should be important for the correct interpretation of strong nonlinearity. For downhill process the disappearance of barrier in torsion energy requires small $(\delta\theta)^2$ (Fig 3, see detailed discussion in section 6 ). In this case, if the frequency difference between initial and final state is large enough then the large $\lambda$ and small $(\delta\theta)^2$ would lead to significant contribution of $\frac{k_B\lambda^2}{2\varepsilon}T^2$ term to $\frac{d\ln W}{d(\frac{1}{T})}$ and therefore the strong non-Arrhenius behavior of temperature dependence can be interpreted. In fact, from Eq(51), the change of slop in a range of temperature $\Delta T$ is

$$\Delta\frac{d\ln W}{d(\frac{1}{T})} = \frac{1}{2}\Delta T + \frac{k_B T\lambda^2}{\varepsilon}\Delta T \qquad (52)$$

By using Eq (47) and (49), the last term of (52) is estimated as

$$\frac{k_B T\lambda^2}{\varepsilon} = k_B T N \frac{\left\langle \ln\frac{\omega_j'}{\omega_j}\right\rangle^2}{\overline{\omega}^2(\delta\theta)^2\langle I_j\rangle} \qquad (53)$$

For example, setting $\delta\theta = 0.1$, $\bar{\omega} = 6 \times 10^{12} \sec^{-1}$, $\langle I_j \rangle = 10^{-37} \text{g} \cdot \text{cm}^2$, $T$=310 K, $\frac{\omega_j'}{\omega_j} \approx$ 2 or 1/2 (1.1 or 1/1.1), and the number of modes $N$ =50, we have $\frac{k_B T \lambda^2}{\varepsilon} = 28(0.53)$. So, as the initial/final frequency ratio larger than 1.1 (or smaller than 1/1.1), the contribution of $\frac{k_B T \lambda^2}{\varepsilon}\Delta T$ term to the slop change will exceed that of $\frac{1}{2}\Delta T$ term. The former contribution increases rapidly with the frequency ratio. As the frequency ratio equals 2 it attains 50 fold higher than the term from $\frac{1}{2}\Delta T$. So, by introducing $\frac{k_B T \lambda^2}{\varepsilon}\Delta T$ term and adjusting the initial/final frequency ratio we are able to interpret all observed non-Arrhenius behaviors in spite of their non-linearity behaving very differently in experiments.

The experimental data to date indicated that most of the ultrafast folders show significant decreases in their folding rates upon increase in temperature at high temperature [Yang and Gruebele, 2004; Ghosh et al, 2007]. The non-Arrhenius behavior has aroused considerable attention of many investigators. It was interpreted by the temperature dependence of hydrophobic interaction or by the nonlinear temperature dependence of the configurational diffusion constant on rough energy landscapes [Scalley and Baker, 1997]. It was also interpreted by elaborately designed conformational searching model [Ghosh et al, 2007]. However, the point of protein folding as quantum transition gives alternative more direct and natural explanation on it.

Recent experiments on rate--temperature relationships in λ-repressor fragments folding exhibited another characteristic of non-Arrhenius behavior. The set of mutants containing more glycines vs alanines exhibit stronger non-Arrhenius behavior in ln$W$ vs 1/$T$ plots [Yang and Gruebele, 2004]. From our point of quantum transition this may be due to the change of torsion potential in the mutation of amino acid. The residue dependence can also be induced by the smaller inertia moment of glycine than alanine since the non-Arrhenius term $\frac{k_B \lambda^2}{2\varepsilon}T^2$ is greater in glycine-rich mutant (see Eq (53)).

As the denaturant concentration effect on $\Delta E$ is considered, the denaturant possibly strengthens the torsion force field and therefore change the slope parameter $S$. $|S|$ will increase with the concentration and therefore the relative contribution of non-Arrhenius term $\frac{\lambda^2}{2\varepsilon}k_B T$ to the rate ln$W$ in Eq (48) will decrease as compared with $S$ term. This interprets why the plots of ln$W$ versus $\frac{1}{T}$ are strongly curved for refolding but almost linear for unfolding [Scalley and Baker, 1997].

## 5 Inertial moment dependence of folding rate

From the simplified equation (26) we obtain the inertial moment dependence of $W$

$$W \sim (\sum I_j)^{-1/2} (\sum \frac{a_j}{I_j})^2 \tag{54}$$

or

$$\ln W \sim -\frac{1}{2} \ln \sum I_j + 2 \ln \sum \frac{a_j}{I_j} \tag{55}$$

The first term at RHS of (55), $\ln \sum I_j$, is called series-connection factor of inertial moment (SCIM) and the second, $\ln \sum \frac{a_j}{I_j}$, the parallel-connection factor of inertial moment (PCIM). Since the folding rates appear to be largely determined by the topology of the native folded state [Plaxco et al, 1998], to calculate two terms of inertial moment Zhang and Luo (2010) proposed a simplified model by calculating the moment of inertia of polypeptide chain between contact residues of native backbone. The calculation is carried out as follows. When the spatial distance between two residues $l$ and $i$ is not greater than some threshold value (for example, 0.8 nm) and the distance in sequence is $i-l>1$, the residue pair is recorded as a contact pair $j(l, i)$. The inertial moment $I_j$ is calculated for each pair $j(l, i)$ and the contribution from all residues between $l$ and $i$ should be summed. For the dataset of 80 proteins collected by Ouyang and Liang (2008) Zhang and Luo calculated the inertial moment dependence of folding rate and found both terms are correlated well with experimental folding rate $\ln k_f$. The correlation coefficient of SCIM with $\ln k_f$ is 0.86 while the correlation coefficient of PCIM with $\ln k_f$ is -0.84. Both correlations are higher than or comparable with other calculations and predictions for protein folding rate based on the same concept of contact order [Plaxco et al, 1998; Ivankov et al, 2003; 2004]. Moreover, they demonstrated the correlation coefficients are insensitive to the choice of the axes for inertial moment calculation. The above results show that the inertial moment as a dynamical variable describing the contact plays an important role in the determination of protein folding rate. They named this approach as dynamical contact order, emphasizing the dynamical aspects of the concept of contact order, distinguished from conventional geometrical and phonological approaches [Segal, 2009; Gromiha et al, 2006].

The summations in (54) and (55) are taken over all contacts of the protein and for each contact the contributions from all residues have been calculated. In the primary approximation, if the differences among residues are neglected, that is, all residues are assumed to contribute equally to the inertial moment (denoted as $I$) then $\ln W$ correlates roughly with the number of residues $N$ in the chain,

$$\ln W \sim \frac{3}{2} \ln N - \frac{5}{2} \ln I \tag{56}$$

This interprets why the chain length is an important factor in the determination of protein folding rate [Galzitskaya et al, 2003].

## 6 Exergonic and endergonic reaction and ultrafast protein folding

We shall compare the theoretical value of protein folding rates $\ln W$ with experimental value $\ln k_f$ in the dataset of 80 proteins. Since data were generally collected under standard experimental conditions [Maxwell et al, 2005], to compare with experiments it is enough to use the theoretical formula Eq (33) at given temperature. To simplify the folding rate computation, we shall neglect the $(\Delta G_{os})^2$ term in Eq (33). Simultaneously, to make the comparison easily established the following two assumptions will be introduced. First, due to our ignorance of the angular quantum number $a_j$ for the time being we assume $a_j = 1$ first and then estimate the error bought about by the assumption. Second, due to lack of the detailed experimental data on $\Delta G_{os}$ for each protein we assume $\Delta G_{os}$ only taking three values roughly, namely $2ck_B T_0$, 0 and $-2ck_B T_0$ where $c$ is a constant to be determined. Since $a_j$ means the square of angular quantum number of low-lying rotational states it is reasonable to assume its value between 1 and 10. So one may use condition

$$\left| \ln W - \ln k_f \right| < 2 \tag{57}$$

as the criterion of consistency between theory and experiment. Using Eq (33) the criterion reads

$$\left| \Delta G_R - \frac{1}{2} \ln \sum_j I_j + 2 \ln \sum_j \frac{1}{I_j} + const - \ln k_f \right| < 2$$

$$( \Delta G_R = \frac{\Delta G_{OS}}{2k_B T_0} ) \tag{58}$$

where $const$ (=constant) is assumed universal for all proteins in dataset.

As checking the validity of the criterion for each protein we find 80 proteins in dataset can be classified into three classes. For 15 proteins the criterion holds as $\Delta G_R = 0$ (denoted as class 3). For other proteins $\Delta G_R$ should be non-vanishing to satisfy the criterion. For 29 proteins (23 two-state proteins and 6 multi-state proteins) the best-fit $\Delta G_R = 8k_B T_0 > 0$, i.e. $c=4$ (denoted as class 1) and for 36 proteins (13 two-state proteins and 23 multi-state proteins) the best-fit $\Delta G_R = -8k_B T_0 < 0$ (class 2). The folding of class 1 proteins is exergonic while the folding of class 2 proteins is endergonic. The class 3 proteins may be exergonic or endergonic but the heat release or absorption is relatively small. The numerical results on the comparison between theoretical and experimental folding rates are listed in Table 1 [Zhang and Luo, 2010]. In theoretical calculations of folding rate three kinds of rotational axis of inertia moment, C1, C2 and C3, have been assumed. C1 means the link of a pair of contact residues *i* and *j* is taken as the axis of rotation. C2 means for residue *k* between a contact pair *i* and *j*, the axis of rotation is defined by a line across *i* or *j*, perpendicular to the plane (*k*, *i*, *j*). C3 means the inertial moment contributed from the *k*–th residue is calculated relative to an axis across the *k*-1-th residue and perpendicular to the link between *k* and *k*-1. From Table 1 we find that the distribution of the deviation between theoretical and experimental value of folding rate has a sharp peak within the range (-2,2) irrespective of the inertial axis choice, indicating that the theory is consistent with experiments [Zhang and Luo, 2010].

In addition to 80-proteins dataset, by use of 27 two-state folding proteins from the "standard" set which was given by Maxwell etc (2005) and Segal (2009), we test whether the consistency condition | ln$W$-ln$k_f$ |< 2 is satisfied. The result shows that all 27 proteins satisfy the condition in C2 rotational axis case and 25 proteins satisfy the condition in C1 and C3 axes cases. The correlation coefficients between ln$W$ and ln$k_f$ are 0.88，0.92 and 0.91 for C1,C2 and C3 respectively.

In above calculations the Eq (33) is used both for two-state and multi-state proteins. Considering there may exist a time delay factor exp(-τ) in non-two--state (section 2）we shall calculate $\ln W - \ln k_f$ by using Eq (40) instead of Eq (33) for multi-state protein folding.

Rigorously speaking, the time delay factor $\exp(-\tau)$ changes from protein to protein. But in the primary calculation we neglect the detailed differences and assume a common value for all multi-state proteins. The time delay $\tau$ can be estimated by two approaches. One is the maximization of the number of proteins that satisfy | ln$W$ -ln$k_f$ |< 2 by varying $\tau$. Another is the maximization of the correlation coefficient between ln$W$ and ln$k_f$ by varying $\tau$. Both calculations show that the maximization occurs at $\tau = 3.5$. These analyses support the view of the unifying mechanism of the two-state and multi-state protein folding. The results of $\ln W - \ln k_f$ for assumed time delay $\tau = 3.5$ are also listed in Table 1.

**Table 1 Distribution of the deviation between theoretical and experimental folding rate**

| $\tau$ | | ln$W$-ln$k_f$ | | | | |
|---|---|---|---|---|---|---|
| | | (-10,-6) | (-6,-2) | (-2,2) | (2,6) | (6,10) |
| 0 | C1 | 3(0) | 10(0) | 54(26) | 12(8) | 1(1) |
| | C2 | 1(0) | 8(0) | 63(27) | 7(7) | 1(1) |
| | C3 | 2(0) | 9(0) | 60(27) | 8(7) | 1(1) |
| 3.5 | C1 | 3(0) | 12(2) | 60(32) | 4(0) | 1(1) |
| | C2 | 1(0) | 10(2) | 68(32) | 1(1) | 0(0) |
| | C3 | 2(0) | 11(2) | 65(32) | 1(0) | 1(1) |

(-10, -6), (-6, -2) etc in the second line of the table indicate the range of ln$W$ -ln$k_f$. The numerical values in the following lines are the number of proteins in the given range (number of multi-state proteins in parenthesis). C1, C2 and C3 correspond to the three rotational axis cases. $\tau$ means the assumed time delay in the delay factor exp(-$\tau$) for multi-state protein folding.

Our methods can predict the exergonic / endergonic property of the folding reaction. By maximization of the number of proteins that satisfy | ln$W$ -ln$k_f$ |< 2 we are able to decide the symbol of $\Delta G_{os}$ for each protein. $\Delta G_{os}$ is the free energy difference between initial and final states. $\Delta G_{os}$>0 means the torsion vibration free energy decreasing in the reaction, $\Delta G_{os}$<0 means the torsion vibration free energy increasing in the reaction. Under the supposition that the torsion vibration contributes the main part of free energy change in protein folding, one predicts that the protein with $\Delta G_{os}$>0 is exergonic and the protein with $\Delta G_{os}$<0 is endergonic. These predictions are basically consistent with experimental data. For example, the nine high-speed folding proteins with free energy change $\Delta G$<0 listed in literature [Kubelka et al, 2004], namely 1E0L, 1ENH, 1L2Y, 1LMB, 1PIN, 1PRB, 1VII, 2A3D, and 2PDD are exactly exergonic proteins ($\Delta G_R$>0) predicted by us. In comparison with experiments we should notice that $\Delta G_R$ means the torsion free energy of initial state minus that of final state while the free energy $\Delta G$ usually denotes the free energy difference between final state and initial state. Simultaneously, we should notice that the experimental free energy $\Delta G$ includes all changes of energy in the reaction in addition to the contribution from torsion vibration.

The kinetic question of how a protein can fold so fast is regarded as one of the grand challenge of the present protein folding theory [Dill et al, 2004]. Many works studied the ultrafast folding of small designed proteins [Qiu et al, 2002; Zhu et al, 2003]. New experimental techniques for direct observation of ultrafast folding were proposed [Neuweiler et al, 2009]. According to above calculations, the main decision factor for ultrafast folding is $\exp\{\frac{\Delta G_{os}}{2k_B T}\}$ in Eq (33). Exergonic proteins have $\Delta G_{os}$>0 which greatly accelerate the folding process. From Eq (31), large $\Delta G_{os}$ means large $\Delta E$ and $\omega_j > \omega_j'$. So, from the point of quantum transition the necessary conditions for ultrafast folding are: first, the folding is exergonic rather than endergonic ; second, the large initial/final energy gap $\Delta E$ in torsion potential and third, the large initial/final frequency ratio of torsion vibration. On the other hand, the high speed folding generally occurs in downhill processes where the small $\delta\theta$ is another factor to increase the folding rate as seen from Eq (33).

The downhill folding was proposed firstly from theory and then demonstrated in experiments [Bryngelson et al, 1995; Sabelko et al, 1999; Garcia-Mira et al, 2002; Huang et al, 2007]. How to understand downhill folding in quantum transition theory? Consider a typical mode with torsion potential of initial $(k=1)$ and final $(k=2)$ state expressed by harmonic potential

$$U_k(\theta) = \frac{1}{2}I\omega_k^2(\theta - \theta_k^{(0)})^2 + E_k \quad (k=1,2) \tag{59}$$

($\delta E = E_1 - E_2 \quad \delta\theta = \theta_1^{(0)} - \theta_2^{(0)}$）

The intersection of $U_1$ and $U_2$ is denoted as $\theta_c$ (Figure 3). By using Eq (59), from the absence of barrier in downhill process, one obtains

$$(\frac{dU_1}{d\theta})_{\theta=\theta_c} = I\omega_1^2(\theta_c - \theta_1^{(0)}) < 0$$
$$(\frac{dU_2}{d\theta})_{\theta=\theta_c} = I\omega_2^2(\theta_c - \theta_2^{(0)}) < 0 \qquad (60)$$

So, the condition for downhill is

$$\theta_c < \theta_1^{(0)} \qquad \theta_c < \theta_2^{(0)} \qquad (61)$$

When initial frequency $\omega_1$ equals final frequency $\omega_2$, $\omega_1 = \omega_2 = \omega$, $\theta_c$ has a simple expression

$$\theta_c = \frac{1}{2}(\theta_1^{(0)} + \theta_2^{(0)}) + \frac{\delta E}{I\omega^2 \delta\theta} \qquad (62)$$

Inserting (62) into (61) one has

$$-\frac{2\delta E}{I\omega^2} < (\delta\theta)^2 < \frac{2\delta E}{I\omega^2} \qquad (63)$$

It means $(\delta\theta)^2$ should be small in downhill process. This is another factor to speed the folding rate.

As Munoz indicated, the absence of barriers to folding is associated with a singular equilibrium behavior and therefore the downhill folding can be identified by investigating the protein unfolding at equilibrium with some experimental techniques [Garcia-Mira et al, 2002]. They observed thermal unfolding process in *E coli* 2-oxoglutarate dehydrogenase multienzyme complex and found that the protein structure melts gradually in the process. The latter was interpreted by the existence of heterogeneous unfolded structures and the arrival from a definite folded state to these structures need different time. In our theory, if the initial state is a heterogeneous ensemble of unfolded structures with a distribution of different frequencies ($\omega_1$) and different minimum potentials ($E_1$) then the refolding of the protein will show highly nonexponential kinetics and exhibit the characteristic of downhill folding.

The protein folding 'speed limit' was widely investigated in recent years. Following the nucleation-condensation mechanism of folding the rate-limiting step is a random, diffusive search for the native tertiary topology [Fersht,1995; Debe et al, 1999]. Along the same line the 'thruway search' model was proposed [Ghosh et al, 2007]. Following Kramer's model of reaction the reactants must diffuse together. The diffusion control in the protein folding reactions was analyzed [Jacob et al, 1997] and the diffusion-limited contact formation in unfolded protein was studied and quantitatively estimated [Hagen et al, 1996]. In the meantime, by studying the folding of engineered $\lambda$-repressors the folding speed limit was estimated as $\sim (2\mu s)^{-1}$ [Yang et al, 2003].

More recently, through single molecule fluorescence techniques, it has been observed that there exist kinetic events of intrachain contact formation in nanosecond range of folding, beyond the fast folding in microsecond range [Neuweiler et al, 2009]. Then, what is the true speed limit for protein folding?

The problem of protein folding 'speed limit' can be easily viewed from quantum uncertainty relation. From the point of folding as a quantum transition between conformational states, the torsion energy levels are distributed with spacing $\hbar\omega_{tor}$ ($\omega_{tor}$, typical torsion frequency about $10^{12}$ sec$^{-1}$) and therefore the lower limit for the time needed by a folding event should be $10^{-12}$ sec due to the quantum uncertainty relation between energy and time ($\Delta e \Delta t \geq \hbar$). Thus, we suggest a new physical speed limit of picoseconds as the limitation for the local and transient fold formation in polypeptide chain.

## 7 Statistical energy landscape approach and torsion-accessible states

As a multi-atom system the conformation of a protein is fully determined by bond lengths $r_{ij}$, bond angles $\chi_{i-j-k}$ and torsion angles $(\theta_1, \theta_2, ... \theta_N)$. Since torsion angles are most easily changed even at room temperature we assume that the conformation is mainly described by a set of torsion coordinates, that is, the conformational potential U is only a function of $(\theta_1, \theta_2, ... \theta_N)$. For a problem in which $r_{ij}$ and $\chi_{i-j-k}$ cannot be fixed, we assume that the potential should be averaged over these variables so that the energy landscape is described by a set of torsion coordinates only. The average is reasonable since the folding does not involve a single microscopic pathway, but rather a statistical energy landscape [Bryngelson et al, 1995]. Through the statistical average of energy landscape and the corresponding redefinition of the potential the number of torsion-accessible states will be largely increased. For example, for a pair of conformational states which are originally inaccessible, it is impossible to reach each other by changing torsion coordinate (or torsion quantum number in quantum theory) only. However, this is possible after the statistical average of energy landscape and redefinition of the form of torsion potential; that is, they have become torsion-accessible. Therefore, the statistical energy approach effectively expands the set of torsion-accessible states.

The torsion potential $U_{tor}(\theta_1, \theta_2, ... \theta_N)$ is generally argument inseparable. Only when Eq (9) is valid the conformational wave function of the system can be expressed as the product of single conformational functions and the quantum transition problem can be solved. The condition Eq (9) can be approximately satisfied through the statistical average of energy landscape. One may average $U_{tor}(\theta_1, \theta_2, ... \theta_N)$ over $N$-1 coordinates with only one angular coordinate $\theta_j$ left as the variable. The resulting potential is denoted as $U_{tor}^{(j)}(\theta_j)$

$$U_{tor}^{(j)}(\theta_j) = \langle U_{tor}(\theta_1, \theta_2, ... \theta_N) \rangle_{\theta_1, ..., \theta_{j-1}, \theta_{j+1}, ..., \theta_N} \tag{64}$$

The summation of $U_{tor}^{(j)}(\theta_j)$ over $j$ gives an approximate expression of $U_{tor}(\theta_1, \theta_2, ... \theta_N)$.

Therefore, the argument-separable torsion potential can be deduced from the statistical energy landscape approach.

The statistical energy landscape approach gives new quantitative insights into the interpretation of experiments and simulations of protein folding thermodynamics and kinetics [Bryngelson et al, 1995]. Above discussions show that even in quantum theory on protein folding the statistical average of energy landscape is indispensably necessary for obtaining a reasonable solution of the complex problem and deducing a robust picture of the folding which experiments have revealed.

After statistical average of energy landscape and expansion of the set of torsion-accessible states there still exist many torsion-inaccessible states. To study the role of torsion-inaccessible states in protein folding we may use master equation formulation as follows. Suppose the set of folding states denoted by F, the torsion-accessible denatured states denoted by U and other torsion-inaccessible unfolded states by $\Phi$. The three kinds of states obey the reaction equations

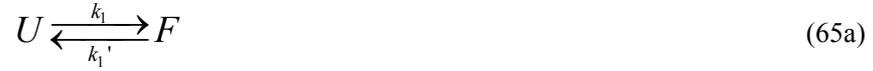
$$U \underset{k_1'}{\overset{k_1}{\rightleftarrows}} F \tag{65a}$$

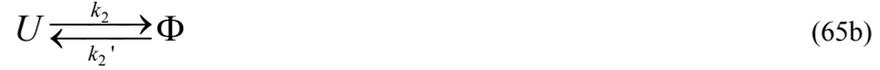
$$U \underset{k_2'}{\overset{k_2}{\rightleftarrows}} \Phi \tag{65b}$$

The direct route between $F$ and $\Phi$ does not exist since the probability of torsion-inaccessible transition has been assumed very small. The rate constants $k_1$ and $k_1$' (as expressed by $W_{AB}$ and $W_{BA}$ in Eq (42)) can be calculated in principle. But we have no enough knowledge on $k_2$ and $k_2$'.

The number of proteins in three sets $N_U$ $N_\Phi$ and $N_F$ changes with time, satisfying the master equation. Due to the existence of pathway (65b) the folding rate from U to F will be influenced by $\Phi$. The rate constants $k_1$, $k_1$', $k_2$ and $k_2$' are dependent of denaturant concentration and temperature. For some concentrations and temperatures the observed protein folding rate will be explicitly changed. The problem will be discussed in further study.

To conclude, the fundamental physics underlying protein folding may be much simpler than the extremely complexity inherent in the protein structure as proposed by David Baker ten years before [Baker, 2000]. Here, our analysis indicates that a key point on the surprising simplicity may be in: the numerous types of protein folding obey the same universal kinetics of quantum transition between conformational states. The proposed theoretical formulation in the article can, in principle, interpret all basic experimental facts on protein folding. Moreover, the view of quantum transition among torsion states gives deeper insights into the folding event of polypeptide chain.

**Acknowledgement**  The author is indebted to Drs Lu Jun and Zhang Ying for their discussions and especially for their help in statistical calculations in section 5 and 6 of this manuscript. Part numerical results in two sections are taken from Zhang and Luo's paper, Acta Scientia Sinica 2010. The author also thanks Dr Chen Wei for his help in the search for several literatures.

# References

Baker D. A surprising simplicity to protein folding. Nature, 2000, 405:39-42.

Bryngelson JD, Onuchic JN, Socci ND, Wolynes PG. Funnels, pathways and the energy landscape of protein folding: A synthesis. Proteins 1995, 21:167-195.

Debe DA, Goddard III WA. First principles prediction of protein folding rates. J Mol. Biol. 1999, 294: 619-625.

Devault Don. Quantum mechanical tunneling in biological systems. Quart. Rev. Biophysics,1980, 13**:** 387-564.

Dill KA，et al. The protein folding problem: when will it be solved? Current Opinion in Structural Biology, 2004, 17:342–346.

Fersht AR. Optimization of rates of protein folding: The nucleation-condensation mechanism and its implications. Proc. Natl. Acad. Sci. USA, 1995, 92**:** 10869-10873.

Fulton KF, Devlin GL, Jodun RA, Silvestri L, Bottomley SP, Fersht AR, Buckle AM. PFD: a database for the investigation of protein folding kinetics and stability. Nucleic Acids Res, 2005, 33:D279CD283.

Galzitskaya OV, Garbuzynskiy SO, Ivankov DN, et al. Chain length is the main determinant of the folding rate for proteins with three-state folding kinetics. Proteins 2003, 51:162-166.

Garcia-Mira MM, Sadqi M, Fischer N, Sanchez-Ruiz JM, Munoz V. Experimental identification of downhill protein folding. Science 2002, 298:2191-2194.

Ghosh K, Ozkan B, Dill KA. The ultimate speed limit to protein folding is conformational searching. J. Am. Chem. Soc. 2007, 129: 11920-11927.

Gromiha MM, Thangakani AM, Selvaraj S. FOLD-RATE: prediction of protein folding rates from amino acid sequence. Nucleic Acids Research 2006, 34(Web Server issue):W70-W74.

Hagen SJ, Hofrichter J, Szabo A, Eaton WA. Diffusion-limited contact formation in cytochrome C: Estimating the maximum rate of protein folding. Proc. Natl. Acad. Sci, USA, 1996, 93: 11615-11617.

Huang F, Sato S, Sharpe TD et al. Distinguishing between cooperative and unimodal downhill protein folding. Proc. Natl. Acad. Sci. USA, 2007, 104: 123-127.

Huang K, Rhys A. Theory of light absorption and non-radiative transitions in F-centers. Proc. Roy. Soc. A , 1950, 204: 406-423.

Ivankov D N, Garbuzynskiy S O, Alm E, et al. Contact order revisited: Influence of protein size on the folding rate. Protein Sci. 2003, 12:2057–2062.

Ivankov D N, Finkelstein A V. Prediction of protein folding rates from the amino acid sequence-predicted secondary structure. Proc. Natl. Acad. Sci. USA, 2004, 101: 8942–8944.

Jacob M, Schindler T, Balbach J, Schmid FX. Diffusion control in an elementary protein folding reaction. Proc. Natl. Acad. Sci. USA, 1997, 94 (11): 5622-5627

Jortner J. Temperature dependent activation energy for electron transfer between biological molecules. J. Chem. Phys. 1976, 64: 4860-4867.

Kakitani T, Kakitani H. A possiblenew mechanism of temperature dependence of electron transfer in photosynthetic systems. Biochim. et Biophys. Acta 1981, 635:498-514.

Kamagata K, Aral M, Kuwajima K. Unification of the folding mechanism of non-two-state and two-state proteins. J Mol. Biol. 2004, 339:951-965.

Kubelka J, Hofrichter J, Eaton W A. The protein folding 'speed limit'. Current Opinion in

Structural Biology, 2004, 14:76–88.

Luo L F. Conformational dynamics of macromolecules. Int J Quant Chem, 1987, 32: 435-450

Luo LF. Conformation transitional rate in protein folding. Int. J. Quant. Chem. 1995, 54: 243-247.

Luo LF. Protein folding as a quantum transition between conformational states. arXiv:0906.2452v1 [q-bio.BM]. 2009, Available from: URL: http://arxiv.org/abs/0906.2452

Luo LF. *Theoretic-Physical Approach to Molecular Biology*. Shanghai : Shanghai Scientific & Technical Publishers, 2004, 388-402.

Maxwell KL, Wildes D, Zarrine-Afsar A, et al. Protein folding: defining a ''standard'' set of experimental conditions and a preliminary kinetic data set of two-state proteins. Protein Sci. 2005, 14(3): 602–616.

Neuweiler H, Johnson CM, Fersht AR. Direct observation of ultrafast folding and denatures state dynamics in single protein molecules. Proc. Natl. Acad. Sci. USA, 2009, 106(44): 18569-18574.

Ouyang Z, Liang J. Predicting protein folding rates from geometric contact and amino acid sequence. Protein Sci. 2008, 17(7):1256–1263.

Pabit SA, Roder H, Hagen SJ. Internal friction controls the speed of protein folding from a compact configuration. Biochemistry 2004, 43:12532-12538.

Plaxco KW, Simons KT, Baker D. Contact order, transition state placement and the refolding rates of single domain proteins. J Mol Biol. 1998, 227(4): 985–994.

Qiu L, Pabit SA, Roitberg AE, et al. Smaller and faster: The 20 residue Trp-cage protein folds in 4 μs. J. Am. Chem. Soc. 2002, 124(44): 12952-12953.

Sabelko J, Ervin J, Gruebelle M. Observation of strange kinetics in protein folding. Proc. Natl .Acad. Sci. USA, 1999, 96: 6031-6036.

Scalley ML, Baker D. Protein folding kinetics exhibit an Arrhenius temperature dependence when corrected for the temperature dependence of protein stability. Proc. Natl. Acad. Sci. USA, 1997, 94: 10636-10640.

Segal M R. A novel topology for representing protein folds. Protein Sci. 2009, 18:686—693.

Yang WY, Gruebele M. Folding at the speed limit. Nature 2003, 423:193-197.

Yang WY, Gruebele M. Rate-temperature relationship in $\lambda$-repressor fragment $\lambda_{6-85}$ folding. Biochemistry 2004, 43:13018-13025.

Zhang Y, Luo LF. The Dynamical Contact Order: Protein folding rate parameters based on quantum conformational transition. SCIENCE CHINA Life Sciences 2010 (to be published).

Zhu Y, Alonso DOV, Maki K et al. Ultrafast folding of $\alpha_3$D : A *de novo* designed three-helix bundle protein. Proc. Natl. Acad. Sci. USA, 2003, 100: 15486-15491.